\documentclass[aps,prb,twocolumn,showpacs,floats,superscriptaddress,preprintnumbers,amsmath,amssymb]{revtex4}

\usepackage{graphicx}
\usepackage{dcolumn}
\usepackage{bm}

\begin{document}

\preprint{(submitted to Phys. Rev. B)}

\title{Water adsorption and dissociation on SrTiO$_{3}$(001) revisited:\\
A density-functional theory study} 

\author{Hannes Guhl}
\affiliation{Fritz-Haber-Institut der Max-Planck-Gesellschaft, Faradayweg 4-6, D-14195 Berlin, Germany}
\affiliation{Institut f\"{u}r Kristallz\"{u}chtung, Max-Born-Str. 2, D-12489 Berlin, Germany}

\author{Wolfram Miller}
\affiliation{Institut f\"{u}r Kristallz\"{u}chtung, Max-Born-Str. 2, D-12489 Berlin, Germany}

\author{Karsten Reuter}
\affiliation{Fritz-Haber-Institut der Max-Planck-Gesellschaft, Faradayweg 4-6, D-14195 Berlin, Germany}
\affiliation{Department Chemie, Technische Universit{\"a}t M{\"u}nchen, Lichtenbergstr. 4, D-85747 Garching (Germany)}


\received{24 February 2010}

\begin{abstract}
We present a comprehensive density-functional theory study addressing the adsorption, dissociation and successive diffusion of water molecules on the two regular terminations of SrTiO$_{3}$(001). Combining the obtained supercell-geometry converged energetics within a first-principles thermodynamics framework we are able to reproduce the experimentally observed hydroxilation of the SrO-termination already at lowest background humidity, whereas the TiO$_2$-termination stays free of water molecules in the regime of low water partial pressures. This different behavior is traced back to the effortless formation of energetically very favorable hydroxyl-pairs on the prior termination. Contrary to the prevalent understanding our calculations indicate that at low coverages also the less water-affine TiO$_2$-termination can readily decompose water, with the often described molecular state only stabilized towards higher coverages.
\end{abstract}

\maketitle

\section{Introduction}

\begin{figure}
\begin{center}
\includegraphics[width = 0.4\textwidth]{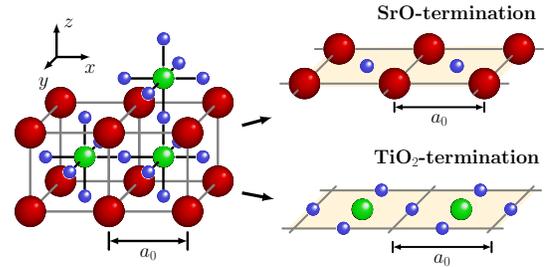}
\end{center}
\caption{(Color online) Schematic illustration of the SrTiO$_3$ perovskite bulk structure and the two regular terminations of the (001)-surface. Big red (dark) spheres: strontium, big green (bright) spheres: titanium, small blue (bright) spheres: oxygen.}
\label{fig1}
\end{figure}

In many of its applications e.g. as photocatalyst, gas sensor or growth template SrTiO$_3$ is deliberately or unintentionally exposed to aqueous environments. This has motivated a large number of studies on fundamental aspects of the interaction of water with this particular material \cite{henderson02}. As for many other SrTiO$_3$ properties, they have most of all revealed a sensitive dependence on the detailed surface morphology and therewith preparation procedure. Within the Surface Science philosophy this shifts the focus to the nominally ``ideal'' two non-polar terminations of SrTiO$_3$(001), cf. Fig. \ref{fig1}, aiming to establish firm answers for these well defined references. The picture emerging from corresponding studies is an enhanced reactivity of water with the SrO-termination\cite{cox83,wang02,owen86,eriksen87}. While this termination seems to hydroxylate readily, weak molecular adsorption appears favored at the TiO$_2$-termination in humid environments. Still, primarily due to the difficulties of preparing a perfect SrTiO$_3$(001) surface that exhibits only one defect-free termination, a discomforting degree of uncertainty concerning the adsorption state, structure and energetics remains. A key experiment in this respect is therefore the Friction Force Microscopy (FFM) study by Iwahori {\em et al.}, which exploits the FFM scanning capability to explore the adsorption of water separately on TiO$_2$- and SrO-terminated domains \cite{iwahori03}. For a range of water exposures a change of the friction force was only measured above SrO-terminated domains and attributed to surface hydroxylation. This confirms the stronger affinity of water to this domain and sets an upper bound for the water bond strength at the TiO$_2$-termination.

Several theoretical works have already been performed to complement these experimental insights. Wang {\em et al.} carried out first density-functional theory calculations (DFT) with a gradient corrected exchange-correlation (xc) functional and determined a binding energy of molecular water at the TiO$_2$-termination in somewhat the range indicated by experiment \cite{wang05}. Unfortunately, the calculations were restricted to small surface unit-cells (thereby modeling a high overlayer density), the TiO$_2$-termination, and did not compare dissociative and molecular adsorption modes. Using a hybrid xc functional Evarestov, Bandura and Alexandrov expanded on the latter two points \cite{evarestov07}. However, in their calculations molecular and dissociative adsorption mode were essentially degenerate at the SrO-termination and only marginally more stable than molecular water at the TiO$_2$-termination -- thereby questioning the prevalent understanding of a more facile surface hydroxylation of the prior termination. Very recently, Baniecki {\em et al.} showed that the similar stability at the two terminations is an artefact of the employed small surface unit-cells \cite{baniecki09}. However, their calculations in larger unit-cells did not systematically compare dissociated and molecular adsorption states.

This situation motivates us to revisit the problem with a comprehensive DFT study that systematically adresses the adsorption, dissociation and decomposition of water molecules at both regular SrTiO$_3$(001) terminations. We show that apart from the coverage-dependence already identified by Baniecki {\em et al.}, also the employed slab thickness is a hitherto not sufficiently appreciated factor for the binding energetics. The finally obtained converged results are in full agreement with the view deduced from experiment, albeit with a small additional twist: Both terminations are able to dissociate water at low coverage without appreciable barrier. The molecular adsorption state discussed at the TiO$_2$-termination is only stabilized at higher coverages. At the SrO-termination the dissociated water stabilizes in hydroxyl pairs, which we identify as the main factor behind the higher water affinity of this termination observed in the FFM measurements.

\section{Theory} 

All DFT calculations were performed with the CASTEP \cite{clark05} code using a plane-wave basis together with ultrasoft pseudopotentials \cite{vanderbilt90} as provided in the default library, and the GGA-PBE functional \cite{perdew96} to treat electronic exchange and correlation. Adsorption at the two regular SrTiO$_3$(001) terminations shown in Fig. \ref{fig1} was modeled in supercell geometries using mirror-symmetric slabs based on the GGA-PBE optimized bulk lattice constant ($a_{0} = 3.938$\,{\AA}), with adsorption at both sides and a vacuum separation exceeding 13\,{\AA}. As further detailed below long-range geometric relaxation effects lead to a slow convergence of the binding energetics with slab thickness. We therefore employed up to 11 layer slabs, in which the central three atomic layers were always fixed at their respective bulk positions in the cubic ($Pm\bar{3}m$) perovskite phase, while all other substrate atoms were fully relaxed until the absolute value of all corresponding forces dropped below 0.05\,eV/{\AA}.

The central energetic quantity taken from the DFT calculations is the binding energy with respect to gas-phase water, defined as
\begin{equation}
\label{eq1}
E_{\rm b} \;=\; \frac{1}{2} \left[ E_{\rm H_2O@surf} - E_{\rm surf} - 2E_{\rm H_2O(gas)} \right] \quad .
\end{equation}
Here $E_{\rm H_2O@surf}$ is the total energy of the adsorbate covered surface (either with molecular or dissociated water), $E_{\rm surf}$ the total energy of the clean surface, and $E_{\rm H_2O(gas)}$ the total energy of the gas-phase H$_2$O molecule (all three computed at the same plane-wave cutoff). The factor $1/2$ accounts for the fact that adsorption is at both sides of the slab and in the sign convention of eq. (\ref{eq1}) a negative value of the binding energy indicates that adsorption is exothermic. The reference energy of the isolated water molecule $E_{\rm H_{2}O(gas)}$ was calculated in a $(20 \times 20 \times 20)$\,{\AA} super-cell resulting in an optimized OH-bond length of 0.99\,{\AA} and a dissociation energy of $-2.47$\,eV as compared to the experimental 0.96\,{\AA} and $-2.52$\,eV, respectively \cite{thiel87}. Systematic tests in $(1 \times 1)$ surface unit-cells indicate that a cut-off energy for the plane-wave basis set of 430\,eV and $(8 \times 8 \times 1)$ Monkhorst-Pack (MP) grids \cite{monkhorst99} for the Brillouin zone integrations ensure a numerical convergence of $E_{\rm b}$ within $\pm 20$\,meV. For calculations in larger surface unit-cells the MP grids were reduced to maintain the same sampling of reciprocal space.

Energetic barriers and minimum energy paths (MEPs) have been computed using the nudged elastic band (NEB) method \cite{henkelman00} in the implementation of the ``Atomic Simulation Environment'' \cite{bahn02}. Corresponding calculations relied on seven atomic layer slabs and $(3 \times 3)$ surface unit-cells throughout. Systematic tests confirm that neither for the obtained transition nor initial and final state geometries spin-polarization plays a role. Furthermore, in all conformations relevant for this work no artificially delocalized electronic states were encountered that could result from the employed semi-local xc functional and which have for example been clearly demonstrated for the case of hydrogen impurities in bulk TiO$_2$.\cite{finazzi08}

\section{Results}

\subsection{SrO-termination}

\begin{figure}
\includegraphics[width = 0.4\textwidth]{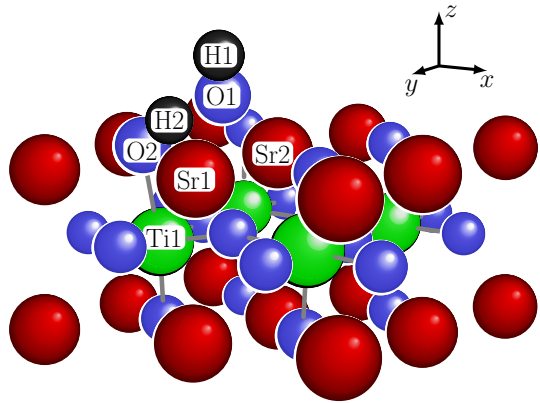}
\caption{
\label{fig2}
(Color online) Perspective view of the most favorable adsorption geometry A1 at the SrO-termination, in which the water dissociates into an adjacent hydroxyl pair. Atoms discussed in the text and Table \ref{tableI} are labeled. Coloring here and in all consecutive figures follows the one of Fig. \ref{fig1}, with small black spheres denoting hydrogen atoms.}
\end{figure}

\begin{table}
\begin{center}
\begin{tabular}{|l|c||l|c|c|}
\hline
\multicolumn{2}{|c||}{SrO-termination} & \multicolumn{3}{c|}{TiO$_2$-termination} \\
\multicolumn{1}{|c}{\quad}  & A1          & \multicolumn{1}{c}{\quad}  & B1          & B2          \\ \hline
H1--O1  & 0.97\,{\AA} & H1--O1  & 0.98\,{\AA} & 0.98\,{\AA} \\ 
H2--O2  & 1.01\,{\AA} & H2--O2  & 1.73\,{\AA} & 0.98\,{\AA} \\
H2--O1  & 1.60\,{\AA} & H2--O1  & 1.01\,{\AA} & 2.78\,{\AA} \\ \hline
O1--Sr1 & 2.59\,{\AA} &         &             &             \\
O1--Sr2 & 2.55\,{\AA} & \raisebox{2ex}[-2ex]{O1--Ti1} & \raisebox{2ex}[-2ex]{2.21\,{\AA}} & \raisebox{2ex}[-2ex]{1.84\,{\AA}} \\ \hline
O2--Ti1 & 2.28\,{\AA} & Ti1--O3 & 1.94\,{\AA} & 2.29\,{\AA} \\ \hline
\end{tabular}
\end{center}
\caption{
\label{tableI}
Selected bond distances in the dissociated adsorption geometry A1 at the SrO-termination, and the molecular B1 and dissociated B2 state at the TiO$_2$-termination. The labels for the individual atoms are defined in Figs. \ref{fig2} and \ref{fig5}.}
\end{table}

Testing all high-symmetry sites offered by the regular SrO-termination we obtain a clear energetic preference for the adsorption geometry A1 displayed in Fig. \ref{fig2} and further quantified in Table \ref{tableI}. In this geometry the water molecule dissociates into two adjacent hydroxyl groups. The O1 and H1 atom from the dissociated water generate a protruding hydroxyl group. Here, the O1 atom sits essentially in a bridge site between two surface strontium cations Sr1 and Sr2, which is the position it would also take in a continuation of the perovskite lattice. The split-off H2 atom of the adsorbing water forms a second hydroxyl group together with a lattice oxygen anion O2. The two hydroxyl groups are strongly tilted towards each other, suggesting the formation of a hydrogen-bond that we will further qualify below. Corresponding bonds with a length comparable to the $d_{\rm O1-H2} = 1.6$\,{\AA} determined here have recently also been reported for hydroxyl adsorption geometries on alkaline-oxide surfaces \cite{carrasco08}.

\begin{table}
\begin{center}
\begin{tabular}{l|ccc}
       & \multicolumn{3}{c}{Surface unit-cell}                               \\
$N_{\rm slab}$ & $(1 \times 1)$              & $(2 \times 2)$ & $(3 \times 3)$    \\ \hline
       & A1                        & A1      & A1            \\
 7     & $-0.83$                   & $-1.18$ & $-1.22$       \\
 9     & $-0.84$                   & $-1.24$ & $-1.30$       \\
11     & $-0.83$                   & $-1.28$ & $-1.35$       \\ \hline
 5     & $-0.79$\cite{evarestov07} & --      & --                \\
 7     & $-0.81$\cite{baniecki09}  & $-1.10$\cite{baniecki09} & -- \\
\end{tabular}
\end{center}
\caption{Binding energies of one water molecule per surface unit-cell in the most favorable adsorption state A1 at the SrO-termination shown in Fig. \ref{fig2} and as a function of the number of slab layers $N_{\rm slab}$ employed. Also shown are the values computed by Evarestov {\em et al.} \cite{evarestov07} using a hybrid xc functional and by Baniecki {\em et al.} using a gradient-corrected xc functional similar to the one employed here. All values in eV.}
\label{tableII}
\end{table}

The hydroxylation lifts the lattice O2 anion quite strongly out of the surface plane. This increases the bond length to the underlying Ti1 atom by 19\,\%. In turn, the latter reinforces its bond to the underlying third layer O anion with a reduction of the corresponding bond length by 25\,\%. This long range geometric relaxation sequence is the major cause for a slow convergence of the binding energy with the number of slab layers employed in the calculations. As detailed in Table \ref{tableII} even for 11 layers slabs there are still remarkable changes of $E_{\rm b}$ of the order of 50\,meV in the larger surface unit-cells. As the geometric relaxations can not properly develop in the smallest $(1 \times 1)$ cells, there is no such dependence there. On the one hand this reveals how misleading slab-thickness convergence tests in such cells can be, on the other hand it allows us to compare our results to those reported in the earlier studies by Evarestov {\em et al.} \cite{evarestov07} and Baniecki {\em et al.} \cite{baniecki09}, which employed corresponding thin slabs in their calculations restricted to smaller cells. As apparent from Table \ref{tableII} (and with equivalent findings compiled in Table \ref{tableIII} for the TiO$_2$-termination) the numbers compare all very well, despite the use of a hybrid xc functional in the study by Evarestov {\em et al.} \cite{evarestov07}. We take this as confirmation that also the gradient-corrected xc functional employed here is able to capture the essential physics -- at a (still) significantly lower computational cost that in turn enables us to perform calculations in much larger surface unit-cells. The data for these cells also compiled in Table \ref{tableII} on the other hand demonstrates that the latter is a crucial point as the binding energy exhibits a strong coverage dependence, reflecting overall repulsive interactions consistent with the TPD data from Wang {\em et al.} \cite{wang02}. A value fairly representing the low-coverage limit is only reached in $(2 \times 2)$ cells, and as we will discuss in Section III.C below, it is this limit that is the appropriate one to discuss the Iwahori FFM experiments \cite{iwahori03}.

\begin{figure}
\begin{center}
\includegraphics[width = 0.4\textwidth]{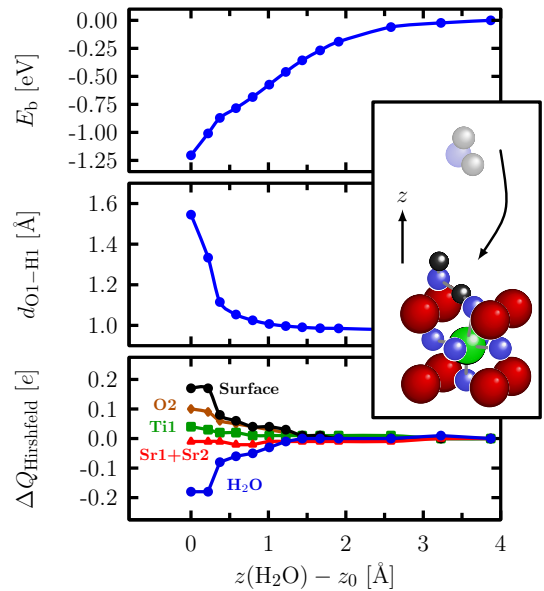} 
\end{center}
\caption{
\label{fig3}
(Color online) Binding energy, internal O1-H2 bond length and adsorption induced change of the Hirshfeld charges as a function of the vertical height $z$ of the center of mass of an approaching water molecule above the SrO-termination. The different atom labels follow the definition given in Fig. \ref{fig2}. The zero reference for the vertical height corresponds to the equilibrium height in the adsorbed state A1.}
\end{figure}

At these relevant low coverages the dissociated water state A1 can be reached without any energetic barriers. Figure \ref{fig3} shows the evolution of the binding energy as a function of the molecule's vertical height above the surface. The barrierless increase of the binding energy upon approaching the surface is accompanied by a gradual charge transfer to the molecule as reflected by the computed Hirshfeld charges \cite{hirshfeld77} also displayed in Fig. \ref{fig3}. The adsorption process fits therefore perfectly into the picture for dissociative adsorption of water, with the water as a Lewis acid and the surface, primarily the O2 anion, acting as a Lewis base \cite{henderson02,thiel87}. 

\begin{figure*}
\begin{center}
\includegraphics[width = 0.90\textwidth]{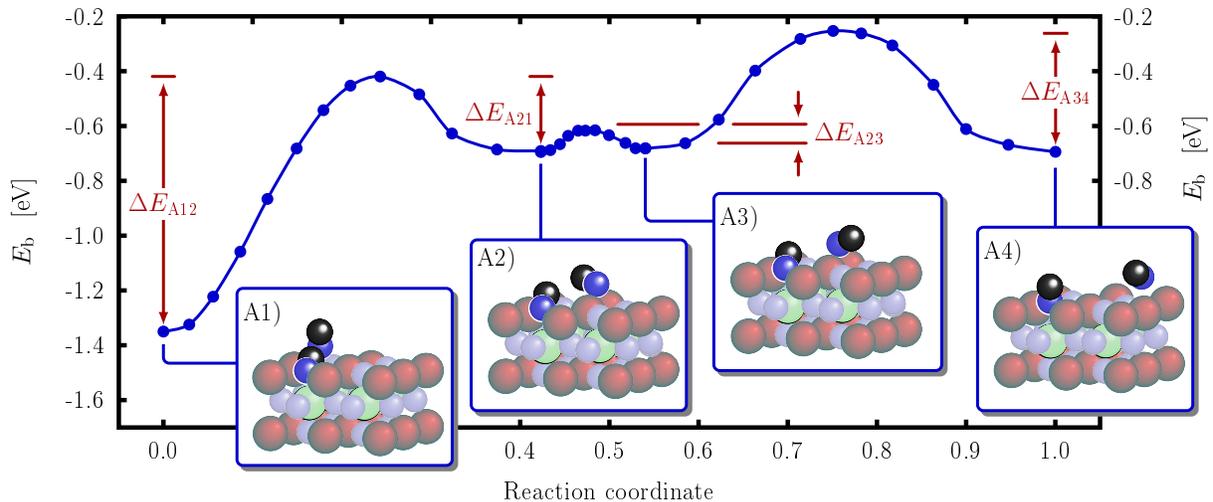}
\end{center}
\caption{\label{fig4}
(Color online) Energy profile for surface diffusion of the protruding O1-H1 hydroxyl group over the SrO-terminated surface. The initial state A1 corresponds to the hydroxyl-pair after the immediate water dissociation process shown in Fig. \ref{fig3}. The transition from state A1 to A2 breaks the hydroxyl-pair and is accompanied by a substantial energetic barrier. Configurations A2 and A3 differ essentially in the mutual orientation of the two hydroxyl-groups, whereas in state A4 the diffusing species has further increased the distance to its original position A1 by hopping to the adjacent binding site.}
\end{figure*}

While the dissociation into the adjacent hydroxyl pair is thus non-activated, this is distinctly different for the consecutive decomposition process. Judging from the stability of each hydroxyl group alone, the initially required breaking of the hydroxyl pair will occur via a diffusion hop of the protruding O1-H1 group. Figure \ref{fig4} shows the computed energy profile for this disintegration pathway. Starting from the most favorable geometry A1 a sizable barrier of $\Delta E_{\rm A12} = 0.93$\,eV needs to be surmounted. This leads to a shallow metastable geometry A2, in which the O1-H1 group sits in a perovskite-type bridge site one unit-cell further away than before. With a computed barrier in excess of 1\,eV for consecutive diffusion of the surface H2 atom, the most favorable pathway for further decomposition proceeds instead through the reorientation of the O1-H1 hydroxyl-group to point away from the original hydroxyl partner. This flip is only connected with a negligibly small barrier $\Delta E_{\rm A23}= 0.08$\,eV, indicating a very fast dangling dynamics. The ensuing diffusion of the separated O1-H1 hydroxyl-group through hops to adjacent binding sites is then characterized by an intermediate barrier $\Delta E_{\rm A34} = 0.41$\,eV. Here, we notice that the energetic minima corresponding to the equivalent states A3 and A4 are aligned on the same level. This suggests no further significant long-range interaction between the immobile O2-H2 group and the moving O1-H1 hydroxyl. On the other hand, this energetic level is located 0.67\,eV above the one of the initial configuration A1, which thus directly reflects the additional bond strength resulting from the hydroxyl pairing. With such strong attraction between the two hydroxyls generated from the dissociation of one water molecule, the percentage of freely diffusing hydroxyl groups or H atoms resulting from this process will be very small up to very high temperatures.

\subsection{TiO$_2$-termination} 

\begin{figure*}
\begin{center}
\begin{minipage}{0.45\textwidth}
\begin{center}
\includegraphics[width =  1.0\textwidth]{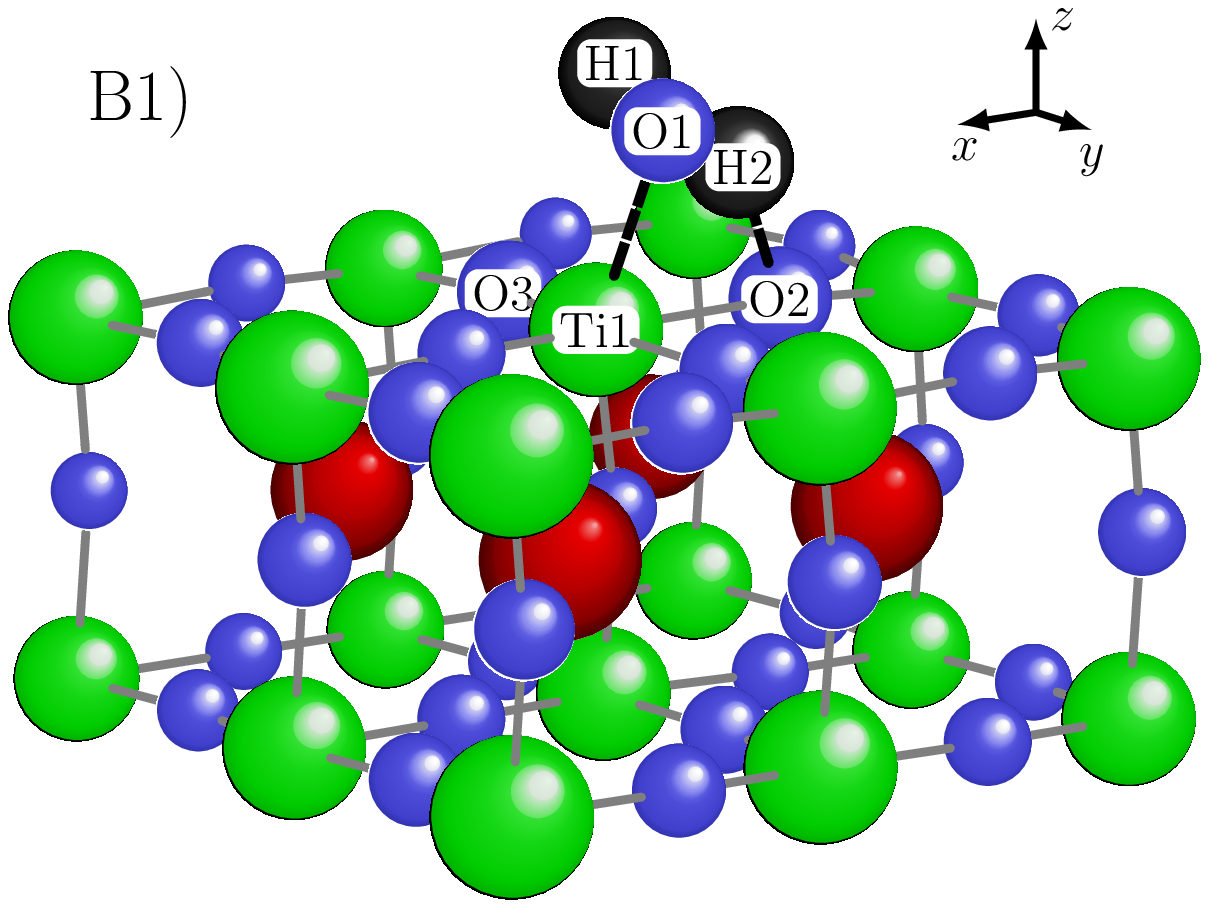}
\end{center}
\end{minipage}
\begin{minipage}{0.45\textwidth}
\begin{center}
\includegraphics[width =  1.0\textwidth]{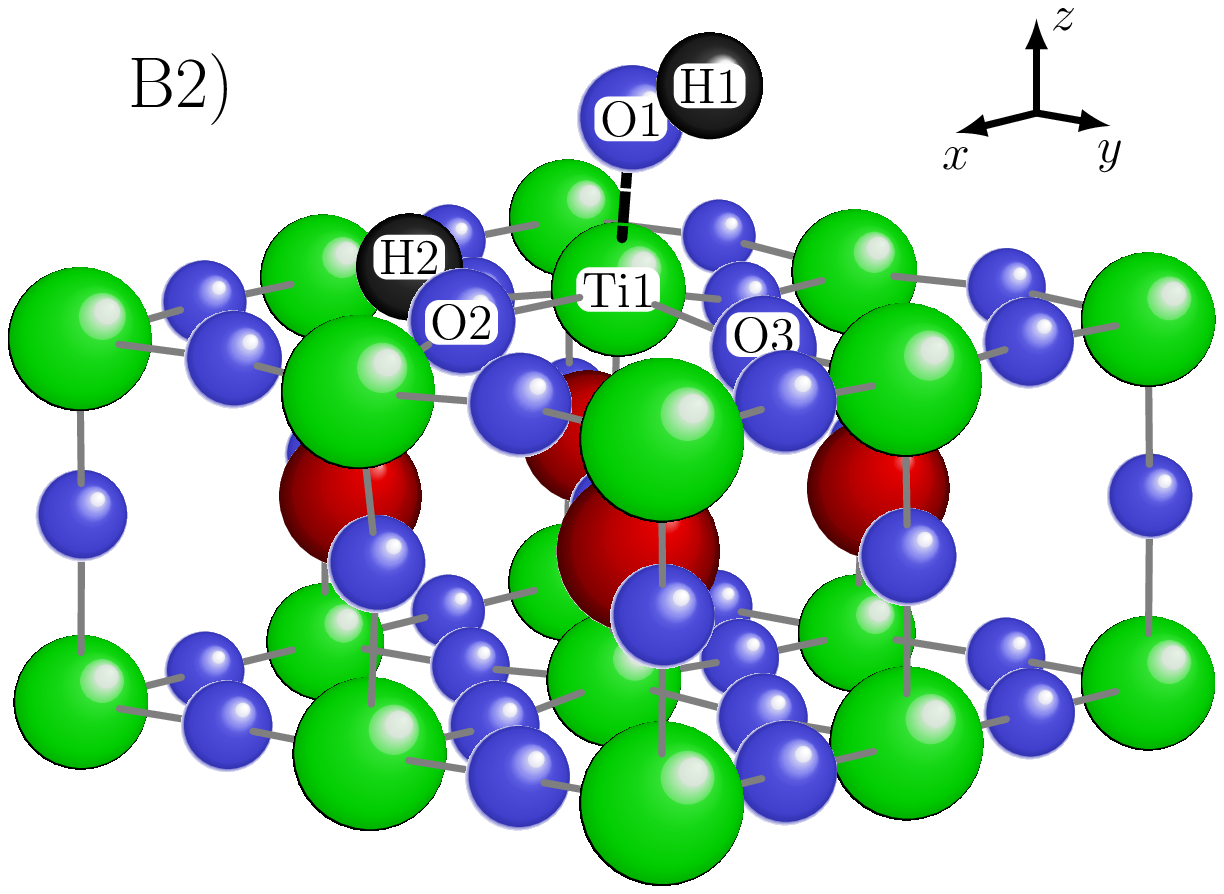}
\end{center}
\end{minipage}
\end{center}
\caption{\label{fig5}
(Color online) Perspective view of the molecular B1 (left) and dissociated B2 (right) adsorption geometry at the TiO$_2$-termination. Atoms discussed in the text and Table \ref{tableI} are labeled. In a second dissociated state (not shown), which is energetically equivalent to the state B2, the O1-H1 group is rotated by $180^{\circ}$ degrees and thus points in the same direction as the tilted O2-H2 hydroxyl.}
\end{figure*}

\begin{table}
\begin{center}
\begin{tabular}{l|cc|cc|cc}
               & \multicolumn{6}{c}{Surface unit-cell}                               \\
$N_{\rm slab}$ & \multicolumn{2}{c}{$(1 \times 1)$} & \multicolumn{2}{c}{$(2 \times 2)$} & \multicolumn{2}{c}{$(3 \times 3)$} \\ \hline
               & B1      & B2                       & B1      & B2                       & B1      & B2      \\
 7             & $-0.77$ & $-0.58$                  & $-0.74$ & $-0.85$                  & $-0.71$ & $-0.90$ \\
 9             & $-0.78$ & $-0.58$                  & $-0.73$ & $-0.92$                  & $-0.72$ & $-0.97$ \\
11             & $-0.78$ & $-0.59$                  & $-0.73$ & $-0.95$                  & $-0.72$ & $-0.99$ \\ \hline

 5             & $-0.73$\cite{evarestov07} & $-0.64$\cite{evarestov07} & -- & --           & --      & --      \\
 7             & $-0.73$\cite{baniecki09}  & --                        & $-0.74$\cite{baniecki09} & -- & --      & --      \\
\end{tabular}
\end{center}
\caption{Binding energies of one water molecule per surface unit-cell in the two most favorable adsorption states at the TiO$_2$-termination shown in Fig. \ref{fig5} and as a function of the number of slab layers $N_{\rm slab}$ employed. Also shown are the values computed by Evarestov {\em et al.} \cite{evarestov07} using a hybrid xc functional and by Baniecki {\em et al.} using a semi-local xc functional comparable to the one employed here. All values in eV.}
\label{tableIII}
\end{table}

In agreement with the preceding theoretical studies \cite{wang05,evarestov07,baniecki09} we also determine a molecular adsorption state on the TiO$_2$-termination. The adsorption geometry of this state henceforth denoted B1 is depicted in Fig. \ref{fig5}, with selected bond distances compiled in Table \ref{tableI}. Adsorption in this mode induces only small geometric relaxations of the SrTiO$_3$(001) substrate, and accordingly we only observe a weak dependence of the computed binding energy with the number of layers employed in the slab model, cf. Table \ref{tableIII}. The coverage dependence is equally weak and in contrast to the dissociated state at the SrO-termination slightly attractive. We also compute a non-activated adsorption pathway into this molecular bound state, which is essentially characterized by an increasing charge depletion around the H2 atom to stabilize the O1-Ti1 molecule-surface bond.

\begin{figure}
\begin{center}
\includegraphics[width=0.4\textwidth]{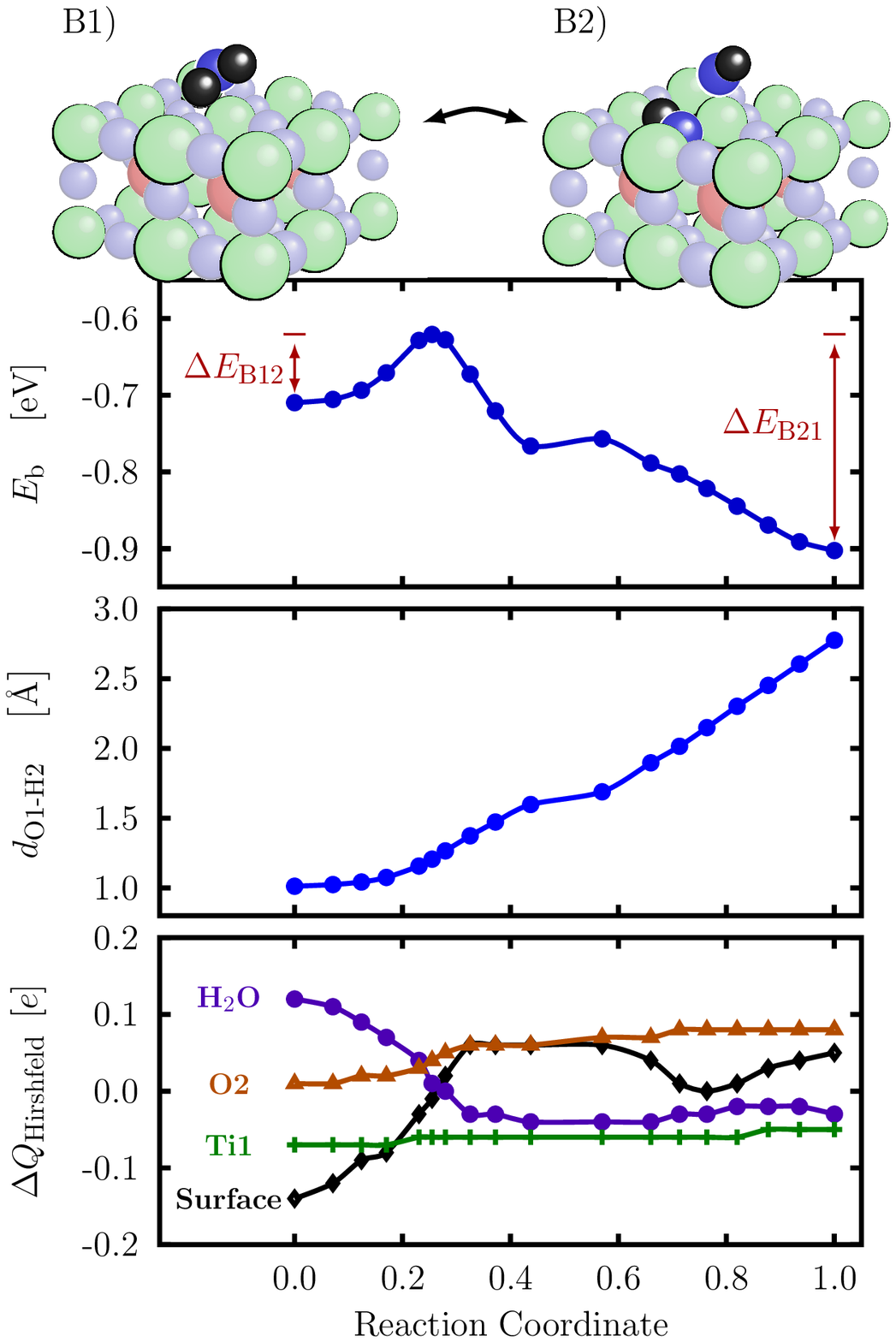} 
\end{center}
\caption{\label{fig6}
(Color online) Binding energy, internal O1-H2 bond length and change of the Hirshfeld charges along the reaction pathway from the molecular bound state B1 to the dissociated state B2 at the TiO$_2$-termination. The different atom labels follow the definition given in Fig. \ref{fig2}. The zero reference for the Hirshfeld charges corresponds to gas-phase water and the clean surface.}
\end{figure}

As illustrated in Fig. \ref{fig6} the full separation of the H2 atom to reach a dissociated state B2 is accompanied by a small energy barrier $\Delta E_{\rm B12} = 0.09$\,eV. In this respect, the molecular state B1 can be considered as a true precursor and its stabilization (in contrast to the SrO-termination) fits well to the expectation of a weaker acidity of the Ti cations compared to the Sr cations on SrTiO$_3$(100) \cite{barr91}. On the other hand, at the low coverage corresponding to the $(3 \times 3)$ surface unit-cell calculation behind Fig. \ref{fig6} the molecular state B1 exhibits only a weakly pronounced metastability, with the ensuing dissociated state B2 (further characterized in Fig. \ref{fig5} and Table \ref{tableI}) more favorable by more than 0.2\,eV. Intriguingly and as detailed in Table \ref{tableIII}, this energetic ordering reverses at higher coverages, with the molecular state B1 more stable in the smallest $(1 \times 1)$ surface unit-cell. We therefore fully reproduce the preference for molecularly adsorbed water obtained in the previous calculations \cite{wang05,evarestov07,baniecki09}, but show that this preference is restricted to rather dense packings. In fact, at these coverages one may well imagine an even further increased stability of mixed phases containing both molecular and dissociated water molecules compared to the two pure phases studied here. We did not pursue calculations along this line though, as we will demonstrate in the following Section that it is the regime of lower coverages that is e.g. relevant for the Iwahori FFM experiments, and in this low coverage regime dissociation into two surface hydroxyl groups is the preferred adsorption mode with only a slight activation barrier to overcome. We note that these findings are in no contradiction to the frequent interpretation of experimental data exclusively in terms of molecularly adsorbed water \cite{cox83,wang02,owen86,eriksen87}. Even though an absolute coverage calibration was rarely achieved, it is quite clear that these studies operated mostly in the higher coverage regime. Moreover and as already discussed by Baniecki {\em et al.} \cite{baniecki09} frequently employed fingerprints for ``molecular'' water (like the appearance of two adsorbate peaks below the O$2p$ valence band in ultraviolet photoemission spectroscopy (UPS) \cite{owen86,eriksen87}) are not necessarily unambiguous. Indeed, we have verified that the valence states due to the hydroxyl groups in the B2 adsorption mode lie at essentially the same energy as the $3a_1$ molecular orbital of the adsorbed water in the B1 mode, such that even a larger fraction of dissociated water would not easily be distinguished in the UPS spectra.

\begin{figure*}
\begin{center}
\includegraphics[width=0.95\textwidth]{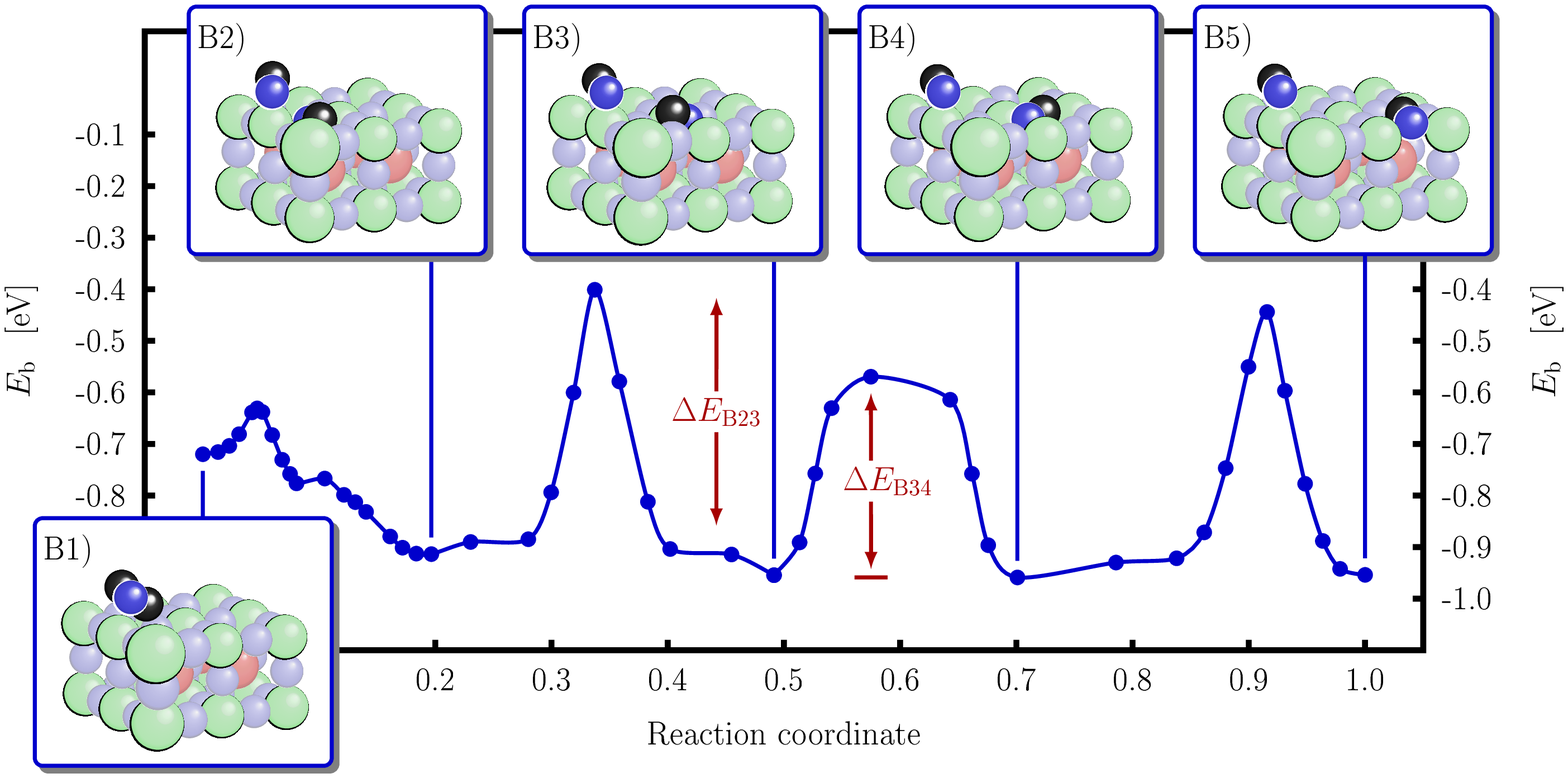}
\end{center}
\caption{\label{fig7}
(Color online) Energy profile for surface diffusion of the split-off H2 atom over the TiO$_2$-terminated surface. The initial states B1 and B2 correspond to the molecular precursor and dissociated hydroxyl-pair shown in Fig. \ref{fig5}. Further disintegration takes place via a hop of the surface H2 atom to an adjacent lattice O anion (B3) with a barrier $\Delta E_{\rm B23} = 0.51$\,eV, followed by a flip of the orientation of the newly formed hydroxyl group towards the neighboring surface unit-cell (B4) with a barrier $\Delta E_{\rm B34} = 0.38$\,eV. The ensuing equivalent hop to an even more distant lattice O anion (B5) exhibits essentially the same energy profile, indicating overall only small longer-range lateral interactions between the hydroxyl groups.}
\end{figure*}

In contrast to the SrO-termination the preferred mechanism for a further disintegration of the adjacent hydroxyl-groups of the dissociated state B2 is not via hopping of the protruding O1-H1 group. We compute a large barrier of 1.5\,eV for a corresponding hop to a neighboring Ti cation. Instead, continued diffusion is at this termination much more effective via surface hopping of the split-off H2 atom. As illustrated in Fig. \ref{fig7} this mechanism proceeds through a sequence of hops with a barrier of 0.51\,eV and reorientation flips with a barrier of 0.38\,eV. In the hops the H2 atom breaks the bond to its directly coordinated lattice O partner and jumps to a directly adjacent lattice O anion. In the consecutive flip the thereby formed hydroxyl group changes its tilt direction to the opposite side, therewith enabling another hop to lattice O anions now belonging to a neighboring surface unit-cell. Interestingly, the rate-limiting step of 0.51\,eV for this diffusion mechanism agrees almost perfectly to the calculated activation barrier for proton diffusion in bulk SrTiO$_3$ \cite{muench99}. A disintegration of the hydroxyls generated through the dissociative adsorption of water is thus not as limiting as at the TiO$_2$-termination and might actually proceed both through surface and sub-surface proton diffusion.

\subsection{First-principles thermodynamics}

In order to directly address the Iwahori FFM experiments we now combine the supercell-converged energetics for adsorbed water at the two terminations and at different coverages within a first-principles atomistic thermodynamics framework \cite{weinert86,scheffler88,kaxiras87,qian88,reuter02a}. In this approach we assume the surface to be in equilibrium with a surrounding water vapor environment characterized by a chemical potential $\mu_{\rm H_2O}$. For each surface termination the stable surface structure at a given $\mu_{\rm H_2O}$ then minimizes the surface free energy, defined as 
\begin{equation}
\gamma(\mu_{\rm H_2O}) = \frac{1}{A} \left[ G_{\rm H_2O@Surf} - G_{\rm Surf} - N_{\rm H_2O} \mu_{\rm H_2O} \right] \quad .
\label{surffreeeng}
\end{equation}
Here, $G_{\rm H_2O@Surf}$ is the Gibbs free energy of the surface covered with $N_{\rm H_2O}$ water molecules per surface unit-cell, $G_{\rm Surf}$ is the Gibbs free energy of the corresponding clean surface, and $\gamma(\mu_{\rm H_2O})$ is normalized to energy per unit area by dividing through the surface area $A$ of the employed surface unit-cell. Aiming only for a first assessment we approximate the difference $(G_{\rm H_2O@Surf} - G_{\rm Surf})$ with the difference of the corresponding total energies, thereby neglecting vibrational free energy and configurational entropy contributions. As discussed in detail in ref. (\onlinecite{reuter02a}) this is largely justified, as most of these contributions effectively cancel in the difference. A notable exception to this arises from the vibrational free contributions from the adsorbate functional groups, which particularly for hydroxyl or water groups are in general not negligible \cite{sun03}. We nevertheless omit them here, realizing from the results presented below that the relevant surface terminations to discuss the Iwahori FFM experiments correspond to rather low adsorbate densities.

Contact to the experimental environments can be made by exploiting the relation between chemical potential and gas-phase temperature and pressure. For this we first separate the total energy contribution to the chemical potential
\begin{displaymath}
\Delta \mu_{\rm H_2O}(T,p_{\rm H_2O}) \;=\; \mu_{\rm H_2O}(T,p_{\rm H_2O}) \;-\; E_{\rm H_2O(gas)} \quad .
\end{displaymath}
At standard pressure $p^{\circ} = 1$\,bar the relative $\Delta \mu_{\rm H_2O}(T,p^{\circ}_{\rm H_2O})$ can then be derived from enthalpy $H$ and entropy $S$ differences tabulated in thermochemical tables \cite{JANAF}
\begin{eqnarray*}
\Delta \mu_{\rm H_2O}(T,p_{\rm H_2O}^{\circ}) &=& \left( H(T,p_{\rm H_2O}^{\circ}) - H(0{\rm K},p_{\rm H_2O}^{\circ} \right) \; - \nonumber \\
                                              && T \;\left( S(T,p_{\rm H_2O}^{\circ}) - S(0{\rm K},p_{\rm H_2O}^{\circ}) \right) \quad .
\end{eqnarray*} 
From this the relative chemical potential at any other pressure follows finally from the ideal-gas relation
\begin{displaymath}
\Delta \mu_{\rm H_2O}(T,p_{\rm H_2O}) \;=\; \Delta \mu_{\rm H_2O}(T,p_{\rm H_2O}^{\circ}) + 
                                            k_{\rm B}T \left(\frac{p_{\rm H_2O}^{\circ}}{p_{\rm H_2O}}\right) \quad ,
\end{displaymath}
where $k_{\rm B}$ is the Boltzmann constant. The latter relation is appropriate as long as the discussion is restricted to humid environments, i.e. those corresponding to the gaseous state of water for any temperature or pressure. This sets an upper bound for $\Delta \mu_{\rm H_2O} = -0.91$\,eV, which corresponds to the chemical potential of water at the (experimental) critical point \cite{sun03}. As we will see in the following the experimental conditions employed in the Iwahori FFM experiments all fall below this H$_2$O-rich limit.

\begin{figure*}
\begin{minipage}{0.46\textwidth}
\includegraphics[width = \textwidth]{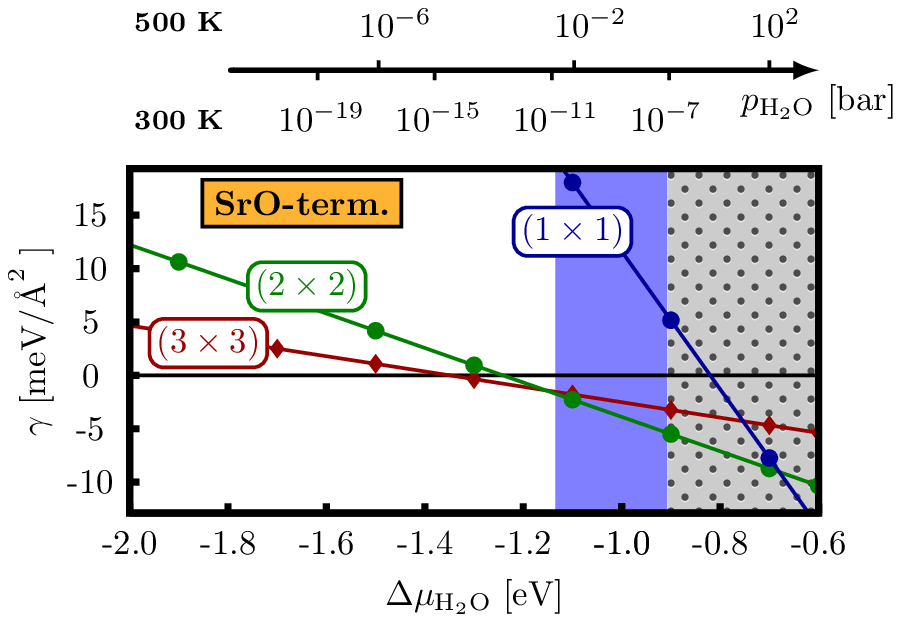}
\end{minipage}
\begin{minipage}{0.46\textwidth}
\includegraphics[width = \textwidth]{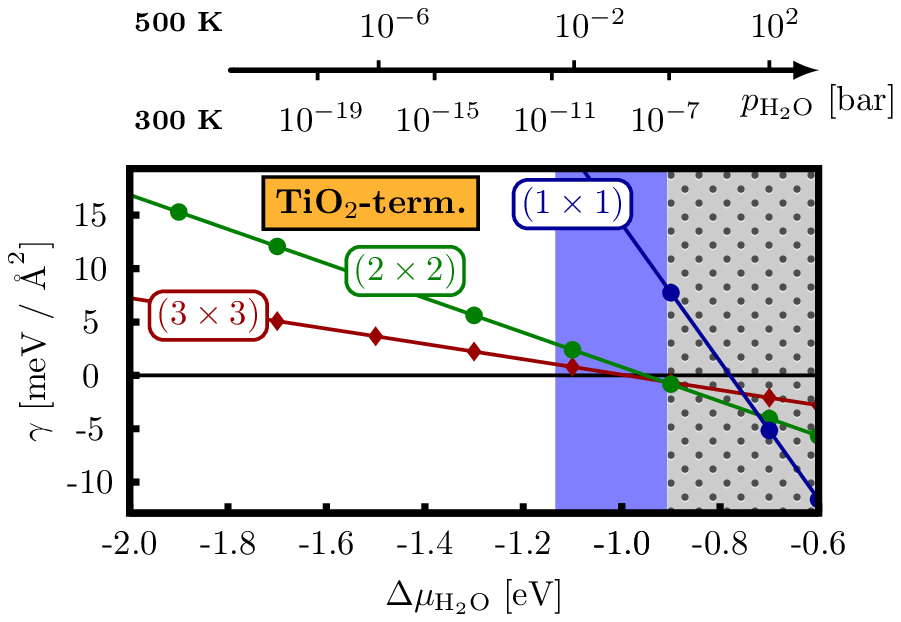}
\end{minipage}
\caption{\label{fig8}
(Color online) Surface free energies, cf. eq. (\ref{surffreeeng}), of the most stable water adsorption geometries at the different coverages at the SrO-termination (left) and TiO$_2$-termination (right). In the top $x$-axis, the dependence on the water chemical potential is converted into pressure scales at 300\,K and 500\,K. The plain (blue) background boxes mark the region of gas-phase conditions probed by Iwahori {\em et al.} in their FFM experiments \cite{iwahori03}, while the dotted (gray) boxes indicate the region above the H$_2$O-rich limit, i.e. where the present approach assuming equilibrium with water vapor is no longer strictly applicable.}
\end{figure*}

Figure \ref{fig8} summarizes the obtained results for the two terminations. Starting the discussion with the SrO-termination, the graph shows the surface free energies obtained for the dissociated hydroxyl-pair state A1 at the three coverages corresponding to one water molecule per $(3 \times 3)$, $(2 \times 2)$ and $(1 \times 1)$ surface unit-cell. Reflecting the absence of significant longer-range lateral interactions between different hydroxyl-pairs at the two lower coverages, the corresponding structures become more stable than the clean surface within a narrow chemical potential range. As apparent from Fig. \ref{fig8} the experimental gas-phase conditions probed in the FFM experiments by Iwahori {\em et al.}\cite{iwahori03} ($10^{-11}$\,bar $< p_{\rm H_2O} < 10^{-7}$\,bar, $T = 300$\,K)  fall clearly above this threshold. The calculations therefore fully support the authors' surmise that the observed increase of the friction-force coefficient on the SrO-terminated domains is due to surface hydroxylation. The observation that the TiO$_2$-termination stays free of water molecules in the same probed environments is also fully consistent with our calculations. The much smaller binding energy for the dissociated state B2 that is most stable at the lower coverages yields a crossing of the corresponding $(3 \times 3)$ and $(2 \times 2)$ lines with the clean surface reference in Fig. \ref{fig8} only at chemical potentials towards the upper end of the experimental range. While our calculations thus confirm the interpretation of the FFM data, the consistency demonstrates vice versa that the supercell-converged absolute binding energetics from the employed semi-local xc functional, which stands behind the thermodynamics lines in Fig. \ref{fig8}, is quite reliable. This concurs with the earlier observation that the semi-local energetics agrees closely to the one obtained by Evarestov {\em et al.} with a hybrid xc functional and reconfirms our assessment that the discrepancies between experiment and theory existing before this work were less due to the treatment of electronic xc, but more due to the use of restricted supercell geometries.

At both terminations the dense $(1 \times 1)$ overlayers that were at the focus of the preceding theoretical works become most stable only at a very high chemical potential of $\Delta \mu_{\rm H_2O} \approx -0.7$\,eV. This is far outside the relevant conditions for the Iwahori experiments and in fact above the H$_2$O-rich limit, where the present thermodynamic approach is no longer strictly valid. This said it is nevertheless intriguing to note that recent FFM experiments by Kato {\em et al.} employed much higher water pressures in excess of $10^{-5}$\,bar at room temperature \cite{kato03}. They initially observed a loss of the Iwahori friction-force contrast between the two SrTiO$_3$(001) terminations, which reappeared at a pressure 
of $10^{-2}$\,bar. The latter threshold corresponds to a $\Delta \mu_{\rm H_2O} = -0.6$\,eV, which is only by about 0.1 eV smaller than where our calculations would predict the stabilization of the dense $(1 \times 1)$ overlayers, and specifically the hydroxylated A1 phase on the SrO-termination and the molecular B1 phase on the TiO$_2$-termination. In contrast, for lower chemical potentials down to $\Delta \mu_{\rm H_2O} = -0.77$\,eV (corresponding to the lower pressure limit of $10^{-5}$\,bar employed by Kato {\em et al.}) both terminations would simply be hydroxylated with less dense arrangements. If hydroxylation is indeed the contrast mechanism behind the FFM observations, this would be fully consistent with the interpretation in terms of the formation of a condensed water layer at the threshold pressure of $10^{-2}$\,bar at room temperature proposed by Kato {\em et al.}\cite{kato03}.

\section{Summary and Conclusions}

In summary our comprehensive DFT calculations demonstrate that both regular terminations of SrTiO$_3$(100) are initially able to dissociate water without appreciable barrier. At the SrO-termination this leads to stable hydroxyl-pairs in the entire sub-monolayer regime. At the TiO$_2$-termination a molecular adsorption state is a metastable precursor at low coverages and becomes the most favorable adsorption mode in a dense overlayer. Combining the computed supercell-converged energetics for these adsorption structures within a first-principles atomistic thermodynamics framework we can fully rationalize the friction-force microscopy experiments by Iwahori {\em et al.}, which indicate a much higher water affinity of the SrO-termination. The absolute binding energetics obtained with the semi-local exchange-correlation functional employed in our study is therefore fully consistent with the stringent bounds set by these measurements. The detailed analysis of the continued disintegration of the dissociated water molecules reveals a strong pairing mechanism of the two surface hydroxyls generated from the dissociation of one water molecule at the SrO-termination. Rather than the stronger acidity of the Sr cations in SrTiO$_3$(001) it is the additional stabilization due to this hydroxyl pairing which stands behind the notably different water affinity of the two terminations. A similar pairing mechanism has recently been reported for the alkaline-oxides CaO and BaO \cite{carrasco08} and we believe this cooperative effect to be an important general feature for low-hydrated oxide surfaces. Another important observation is that the O atom of the protruding hydroxyl group that has formed as a result of the water dissociation process sits in both terminations at the site it would also take in a continuation of the perovskite lattice structure. This is distinctly different to adsorbed O atoms, which we previously found to adsorb in non-perovskite sites \cite{guhl10}. This difference as well as the low mobility of paired hydroxyl groups could be important ingredients towards an atomic-scale understanding of the experimental reports that hydrogen and water increase the growth rate of SrTiO$_3$.\cite{simpson94}

\section*{Acknowledgments}

All calculations have been performed on the SGI Altix ICE1-, ICE2-, and XE- computing clusters of the North-German Supercomputing Alliance (HLRN) 
hosted at the Konrad-Zuse-Zentrum f\"{u}r Informationstechnik in Berlin (ZIB) and at the Regionales Rechenzentrum f\"{u}r Niedersachsen (RRZN) at the Leibniz Universit\"{a}t Hannover. We are particularly grateful to B. Kallies for technical support.


\begin{thebibliography}{46}

\bibitem{henderson02}
M.A. Henderson, Surf. Sci. Rep. {\bf 46}, 1 (2002).

\bibitem{cox83}
P.A. Cox, R.G. Egdell, and P.D. Naylor, J. Electron Spectrosc. Relat. Phenom. {\bf 29}, 247 (1983).

\bibitem{wang02}
L.-Q. Wang, K.F. Ferris, and G.S. Herman, J. Vac. Sci. Technol. A {\bf 20}, 239 (2002).

\bibitem{owen86}
I.W. Owen, N.B. Brookes, C.H. Richardson, D.R. Warburton, F.M. Quinn, D. Norman, and G. Thornton,
Surf. Sci. {\bf 178}, 897 (1986).

\bibitem{eriksen87}
S. Eriksen, P.D. Naylor, and R.G. Egdell, Spectrochim. Acta A {\bf 43}, 1535 (1987).

\bibitem{iwahori03}
K. Iwahori, S. Watanabe, M. Kawai, K. Kobayashi, H. Yamada, and K. Matsushige,
J. Appl. Phys. {\bf 93}, 3223 (2003).

\bibitem{wang05}
L.-Q. Wang, K.F. Ferris, S. Azad, and M.H. Engelhard, J. Phys. Chem. B {\bf 109}, 4507 (2005).

\bibitem{evarestov07}
R.A. Evarestov, A.V. Bandura, and V.E. Alexandrov, Surf. Sci. {\bf 601}, 1844 (2007).
The numbers cited in Tables \ref{tableII} and \ref{tableIII} include a correction of 0.12\,eV
for the basis set superposition error as estimated by the authors.

\bibitem{baniecki09}
J.D. Baniecki, M. Ishii, K. Kurihara, K. Yamanaka, T. Yano, K. Shinozaki, T. Imada, and Y. Kobayashi,
J. Appl. Phys. {\bf 106}, 054109 (2009).

\bibitem{clark05}
S. Clark, M.D. Segall, C.J. Pickard, P.J. Hasnip, M.I.J. Probert, K. Refson, and M.C. Payne, 
Z. Kristallogr. \textbf{220}, 567 (2005).

\bibitem{vanderbilt90}
D. Vanderbilt, Phys. Rev. B \textbf{41}, 7892 (1990).

\bibitem{perdew96}
J.P. Perdew, K. Burke, and M. Ernzerhof, Phys. Rev. Lett. {\bf 77}, 3865 (1996).

\bibitem{thiel87}
P.A. Thiel and T.E. Madey, Surf. Sci. Rep. {\bf 7}, 211 (1987).

\bibitem{monkhorst99}
H. Monkhorst and J. Pack, Phys. Rev. B \textbf{13}, 5188 (1976).

\bibitem{henkelman00}
B.P. Henkelman, G. Uberuaga, and H. Jonsson, J. Chem. Phys. {\bf 113}, 9901 (2000).

\bibitem{bahn02}
S.R. Bahn and K.W. Jacobsen, Comput. Sci. Eng. {\bf 4}, 56 (2002).

\bibitem{finazzi08}
E. Finazzi, C. DiValentin, G. Pacchioni, and A. Selloni, J. Chem. Phys. {\bf 129}, 154113 (2008).

\bibitem{carrasco08}
J. Carrasco, N. Lopez, and F. Illas, Phys. Rev. Lett. {\bf 100}, 016101 (2008).

\bibitem{hirshfeld77}
F.L. Hirshfeld, Theoret. Chim. Acta {\bf 44}, 129 (1977).

\bibitem{barr91}
T.L. Barr, J. Vac. Sci. Technol. A {\bf 9}, 1793 (1991); T.L. Barr and C.R. Brundle, Phys. Rev. B {\bf 46}, 9199 (1992).

\bibitem{muench99}
W. Muench, K.-D. Kreuer, G. Seifertli, and J. Majer, Solid State Ionics {\bf 125}, 39 (1999).

\bibitem{weinert86}
C.M. Weinert and M. Scheffler, In: {\em Defects in Semiconductors},
H.J. von Bardeleben (Ed.), Mat. Sci. Forum {\bf 10-12}, 25 (1986).

\bibitem{scheffler88}
M. Scheffler, In: {\em Physics of Solid Surfaces - 1987}, J. Koukal (Ed.), Elsevier, Amsterdam (1988); 
M. Scheffler and J. Dabrowski, Phil. Mag. A {\bf 58}, 107 (1988).

\bibitem{kaxiras87}
E. Kaxiras, Y. Bar-Yam, J.D. Joannopoulos, and K.C. Pandey, Phys. Rev. B {\bf 35}, 9625 (1987).

\bibitem{qian88}
G.-X. Qian, R.M. Martin, and D.J. Chadi, Phys. Rev. B {\bf 38}, 7649 (1988).

\bibitem{reuter02a}
K. Reuter and M. Scheffler, Phys. Rev. B {\bf 65}, 035406 (2001).

\bibitem{sun03}
Q. Sun, K. Reuter, and M. Scheffler, Phys. Rev. B {\bf 67}, 205424 (2003).

\bibitem{JANAF}
D.R. Stull and H. Prophet, {\em  JANAF Thermochemical Tables}, 2nd ed.,
U.S. National Bureau of Standards, Washington, D.C. (1971).

\bibitem{kato03}
H.S. Kato, S. Shiraki, M. Nantoh, and M. Kawai, Surf. Sci. {\bf 544}, L722 (2003).

\bibitem{guhl10}
H. Guhl, W. Miller, and K. Reuter, Surf. Sci. {\bf 604}, 372 (2010).

\bibitem{simpson94}
T.W. Simpson, J.C. Mitchell, I.V. McCallum, and L.A. Boatner, J. Appl. Phys. {\bf 76}, 2711 (1994).

\end{thebibliography}
\end{document}